\documentclass[11pt,twoside]{article}
\usepackage{asp2014}

\aspSuppressVolSlug
\resetcounters

\begin{document}

\title{When a couple's squabbling leads to cosmic bubbles}
\author{Henri M.J. Boffin}
\affil{ESO, Santiago, Chile\\ \email{hboffin@eso.org}}

\paperauthor{Boffin}{hboffin@eso.org}{}{ESO}{Vitacurat}{Santiago}{}{}{Chile}

\begin{abstract}
Binarity and mass transfer appear to play a key role in the shaping and, possibly, formation of planetary nebulae (PNe), thereby explaining the large fraction of axisymmetric morphologies. I present the binary hypothesis for PNe and its current status. Recent discoveries have led to a quadrupling of the number of post-common envelope binary central stars of PNe compared to the end of the last century, thereby allowing us to envisage statistical studies. Moreover, these binary systems let us  study in detail the mass transfer episodes before and after the common envelope, and I present the evidences for mass transfer -- and accretion -- prior to the common envelope phase. 
\end{abstract}

\section{The binary hypothesis}
Planetary nebulae (PNe) -- these magnificent cosmic bubbles that light up the Universe in a colourful way \citep{Popiplan} -- are generally considered to be the swan song of low- and intermediate-mass (i.e. 1 to 8 M$_\odot$) stars, before  becoming white dwarfs. As such, the central stars of PNe (CSPNe) are the precursors of white dwarfs and their study provides us with the initial conditions to study the latter. They therefore fully deserve to be part of a workshop dedicated to white dwarfs.

The classical scenario to form a PN is as follow: at the end of the Asymptotic Giant Branch (AGB), the star has lost most of its envelope (and thus its mass) through a slow but fierce wind and the hot central star is being exposed. The latter is also undergoing mass loss through a fast, hot wind, which drags along the left-over material of the AGB envelope in a narrow shell, which is then ionised by the hot star. This scenario -- known as the Interacting Wind model \citep{Kwok78} -- should produce a spherical PN. And indeed, the original and misleading name of this class of object, \emph{planetary} nebula, arose because of the roundish aspect of the earliest members discovered. However, with better telescopes, it has soon become clear that a great majority of PNe are not spherical. Many are ellipticals or bipolars, with some presenting clear jet-like structures and/or ``ansae'', while others even appear multipolars \citep{BF02}. As such, planetary nebulae with their intricate and vivid shapes are amazing treats for the eye, but also treasures for the mind, as one need to explain how a spherical star leads to such non-spherical nebulae.
This is a critical question, whose corollary can be formulated as \emph{``are we sure that all (or at least most) white dwarfs go through a PN stage or is there a need to have something else to make a PN?''}

I am of the opinion that, after several decades of debate, it is now safe to assert that the origin of the non-spherical, axisymmetric morphology of PNe lies in the presence of a (sub-)stellar companion \citep{Soker97,NB06,DeM09}. This has led to the so-called  \emph{binary hypothesis} for PNe \citep[][see also Paczynski 1985]{MM06,DeM09}. Its weak variant asserts that ``PNe derive from binary progenitors more easily than from single progenitor'', where ``binary'' should be understood here in a general term, i.e. the central star may have a star, a brown dwarf or even a giant planet as a companion. 

The number of non-spherical PNe amounts to at least 80\% \citep{BF02}.
Can we explain such a high fraction? 
Recent surveys \citep[e.g.,][]{Raga10} indicate that the fraction of binaries among solar-like stars is about 50\%, but given that one should ignore all systems with a separation greater than, say, 200 AU (as these are likely not really interacting binaries), this fraction becomes 30\%. This value includes brown dwarfs. The same authors find that $9\pm2$\% of single stars have a planet, so that in total, one can account for $\sim$40\% of all stars to have at least a companion, stellar or substellar, with which to interact. In addition, one should note that even if the object now appears single, it does not mean it was not a binary at first, as some objects will go through a merger phase \citep{Web76,2014ApJ...786...39N}. Thus, from these arguments alone, it appears that one can explain the binary (and, by extension, the non-spherical morphology) of at least 40\% of all PNe. The fact that we \emph{apparently} see a fraction of such aspherical PNe twice as large may be due to the existence of a population of  ``hidden'', low-surface brightness, spherical PNe \citep{Soker97}. The fraction of these hidden PNe should be between 0 and 50\% of all PNe, depending on the number of mergers, in order not to need the binary hypothesis. Whether such a population exist is still not clear and no conclusion can yet be drawn at this time, except for the fact that the morphology of many PNe is clearly influenced by a companion.
This latter point is certainly proven by the fact that there is now a handful of PNe with binary central stars for which it was possible 
to reliably compare the inclinations of both binary orbital plane and nebular symmetry
axis, and in all cases, these two angles were similar \citep{DJo}. The conclusion is inescapably that the central binary is the cause of the shaping of the nebula. 

\section{Post-CE systems}
An AGB star in a binary system will, unless the system is really very wide, interact in some way with its companion. It will transfer mass by wind \citep{BoffinBSS} and, if the orbital period is smaller than some limiting value, it will overflow its Roche lobe, a process that will generally lead to a common envelope and a spiral-in of the companion \citep{IvanovaBSS}. The final orbit will be very tight, with an orbital period in the range of a few hours to a few days. In this case, the envelope which is ejected is the nebula that will be ionised by the white dwarf. \cite{Mis09} found that the close binary fraction, i.e. the faction of binary CSPNe with orbital period below about 3 days, is about 12--21\%. This value may be slightly underestimating the real close binary fraction due to observational biases, as the photometric method they use will only pick binaries (recognised in the light curves through eclipses, irradiation or elongation effects) if the signal is strong enough, which depends on the time sampling, the inclination of the system, the relative brightness of the nebula and the orbital period. It should, however, provide a good first estimate. 

One can do a rough calculations to see if such a value is compatible with what we know about binary stars. If we take as a rough estimate that the maximum radius an AGB star can reach is about 1000 R$_\odot$, we can  estimate that all systems with a period below about 21--30 years ($\log P(d) \sim 3.89-4.04$) -- depending on the total mass of the system and the mass ratio -- will lead to a common envelope and thus to a short period binary system. \cite{DM91} determined the distribution of orbital periods for solar-like binaries and found that it could be approximated as a gaussian (in logarithm of the period when expressed in days), centred around 4.8 and with a standard deviation of 2.3. The more recent study of  \cite{Raga10} finds similar results, i.e. a gaussian, centred around 5.03 and with a standard deviation of 2.28. Using these last numbers, this implies that about 30 to 32\% of all solar-like binary systems would go through a common envelope phase. Taking into account the 50\% fraction of binaries for solar-like stars, we obtain a close binary fraction of 15--16\%, not unlike what is observed. 
This may also be a lower limit, however, as systems with initially longer periods may also shrink during the wind mass transfer episodes that will precede \citep{BoffinBSS}.
Of course, this back-of-the-envelope calculation should be confirmed by detailed population synthesis estimates, but it already provides some good reason to think that the close binary fraction does not require the PN binary hypothesis to be valid. This, is turn, would imply the existence of a huge population of hidden, spherical PNe. In this respect, it is useful to note that \cite{Schreiber08} found that 35\% of all white dwarf-main sequence (WD-MS) binaries are post-CE systems, in agreement with the rough calculation made above.

\articlefigure{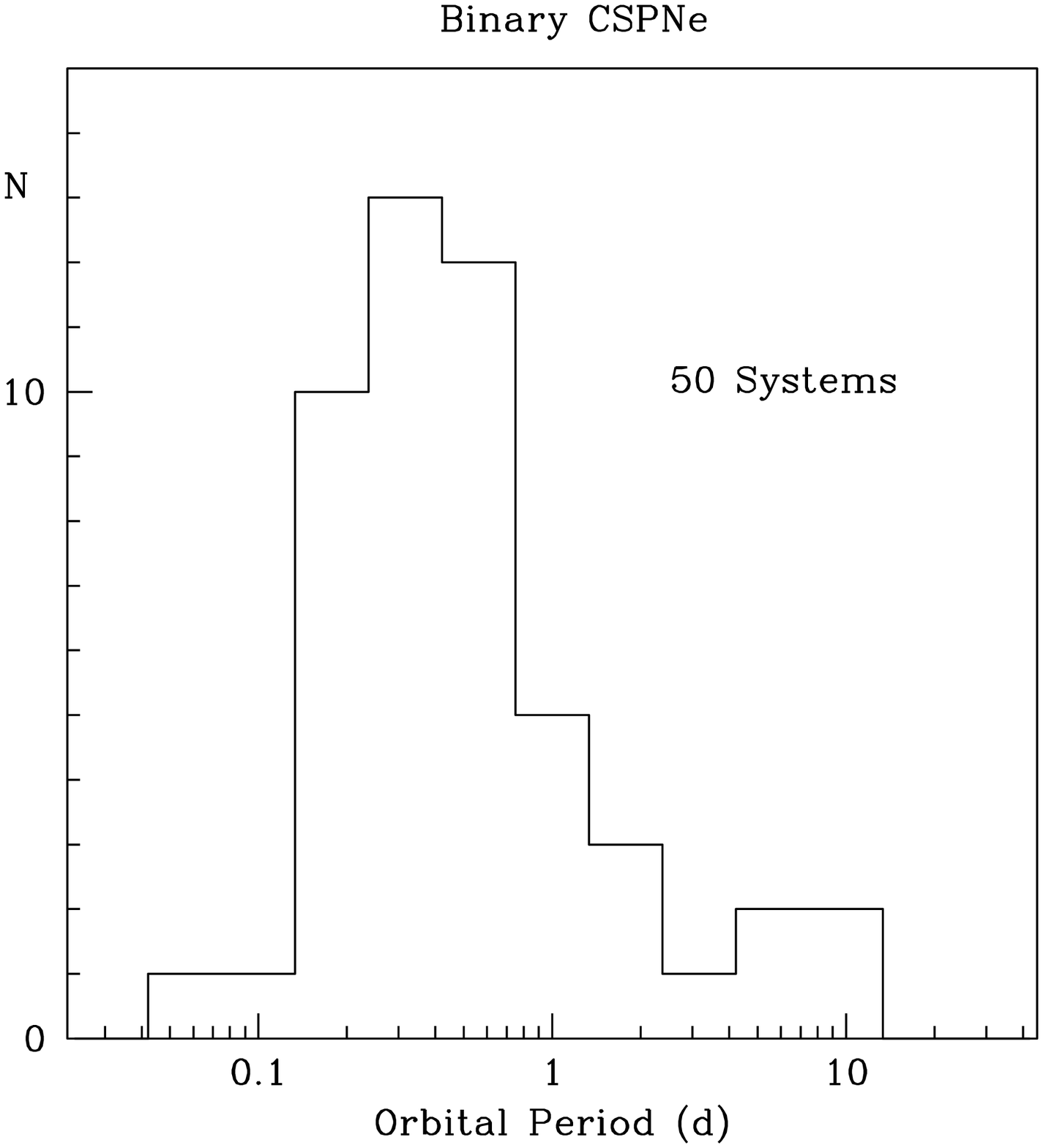}{HB_Fig1}{Distribution of the orbital period (in days) for all known post-CE CSPNe.}

\section{Fresh out of the oven}
In the last few years, the number of close binary CSPNe (i.e. post-CE) has dramatically increased from the dozen known at the turn of the 21st century \citep{Bond00}. Dave Jones is collecting the list of all planetary nebulae with binary central stars\footnote{http://www.drdjones.net/?q=bCSPN} and at the time of writing there are now 50 post-CE PNe known and for which an orbital period has been determined. Thus, in addition to providing useful information on the role of binarity in the shaping and, possibly, formation of PNe, these binary CSPNe can be very useful to study binary populations in general, as binary CSPNe can be thought of as "fresh out of the oven". This is particularly true for post-CE systems and their study should increase our knowledge on the poorly-constrained common envelope process. In particular, the distribution of orbital periods of post-CE CSPNe (Fig.~\ref{HB_Fig1}) should be a necessary tool to constrain models. The availability of 50 systems allows already to have a sizeable sample that can be compared to the outcome of binary populations as well as to other post-CE samples. In this regard, it is noteworthy that the observed period distribution drops dramatically for orbital periods above 1 day. This could partly be due to observational biases, but could as well indicate that the energy efficiency parameter, $\alpha_{\rm CE}$, should be much smaller than unity \citep{Nie12}. The difference with the sample of WD+MS from the SDSS survey \citep{Reb12} may also indicate some fundamental issues in the formation of post-CE PNe.

\section{Proving accretion}
Post-CE PNe allow us also to study the process of mass transfer. The presence of collimated outflows or jets surrounding several systems \citep[e.g.,][]{Bo12}  possibly launched from  an accretion disc that is no longer present is most likely indirect evidence for accretion, either prior to the common envelope phase via wind accretion from the AGB primary, during the start of the CE infall phase or perhaps even after the CE phase.  
Observations indicate the jets were probably ejected before the main nebula \citep{DJo}. Another fundamental clue comes from point-symmetric outflows of PNe. Simulations can recreate these complex outflows with a precessing accretion disc around the secondary launching jets, with Fleming 1 \citep{Bo12} being certainly the prototype. The characteristics of Fleming 1 indicate that wind accretion (with the formation of a disc) must have happened \emph{before} a CE episode.
More convincing still is the detection by \cite{2013MNRAS.428L..39M}  of a carbon dwarf secondary in the post-CE central star binary in the Necklace planetary nebula. To reproduce the observed carbon enhancement, one need to accrete between 0.03 and 0.35~M$_\odot$, depending on the mass of the star, in a binary system with initial orbital period between 500 and 2000 days. The current period of the system, 1.16 d, clearly proves that the system underwent a common-envelope phase. 
The most advanced simulations of the spiral-in part of the CE phase predict a negligible amount of mass accretion, $10^{-3}$~M$_\odot$ \citep[see references in][]{BoffinBSS}. The accretion is therefore most likely to have occurred before the CE phase via wind accretion, a process that simulations predict to form an accretion disc around the companion. Since the jets of the Necklace are also observed to be older than the main PN, they were probably launched from such a disc.



\end{document}